\documentclass[twocolumn,amsmath,amssymb,nofootinbib,prl,superscriptaddress]{revtex4-1}

\usepackage{amsmath,amssymb,array,calc,epsfig}
\usepackage{graphicx}
\usepackage{psfrag,verbatim,bm}
\usepackage{hyperref}
\usepackage{amssymb,graphicx}
\usepackage[usenames]{color}

\begin{document}

\def\ie{{\it i.e.}}
\def\eg{{\it e.g.}}

\def\cala         {{\cal A}}
\def\calA         {{\mathfrak A}}
\def\calAbar      {{\underline \calA}}
\def\calb         {{\cal B}}
\def\calc         {{\cal C}}
\def\cald         {{\cal D}}
\def\cale         {{\cal E}}
\def\calf         {{\cal F}}
\def\calg         {{\cal G}}
\def\calG         {{\mathfrak G}}
\def\calh         {{\cal H}}
\def\cali         {{\cal I}}
\def\calj         {{\cal J}}
\def\calk         {{\cal K}}
\def\call         {{\cal L}}
\def\calm         {{\cal M}}
\def\caln         {{\cal N}}
\def\calo         {{\cal O}}
\def\calp         {{\cal P}}
\def\calq         {{\cal Q}}
\def\calr         {{\cal R}}
\def\cals         {{\cal S}}
\def\calt         {{\cal T}}
\def\calu         {{\cal U}}
\def\calv         {{\cal V}}
\def\calw         {{\cal W}}
\def\calz         {{\cal Z}}

\def\complex      {{\mathbb C}}
\def\naturals     {{\mathbb N}}
\def\projective   {{\mathbb P}}
\def\rationals    {{\mathbb Q}}
\def\reals        {{\mathbb R}}
\def\zet          {{\mathbb Z}}

\def\del          {\partial}
\def\delbar       {\bar\partial}
\def\ee           {{\rm e}}
\def\ii           {{\rm i}}
\def\chain        {{\circ}}
\def\tr           {\mathop{\rm Tr}}
\def\Re           {{\rm Re\hskip0.1em}}
\def\Im           {{\rm Im\hskip0.1em}}
\def\id           {{\it id}}

\def\de#1#2{{\rm d}^{#1}\!#2\,}
\def\De#1{{\cald}#1\,}

\def\half{{\frac12}}
\newcommand\topa[2]{\genfrac{}{}{0pt}{2}{\scriptstyle #1}{\scriptstyle #2}}
\def\undertilde#1{{\vphantom#1\smash{\underset{\widetilde{\hphantom{\displaystyle#1}}}{#1}}}}
\def\prodprime{\mathop{{\prod}'}}
\def\gsq#1#2{%
    {\scriptstyle #1}\square\limits_{\scriptstyle #2}{\,}} % Ginsparg square
\def\sqr#1#2{{\vcenter{\vbox{\hrule height.#2pt
 \hbox{\vrule width.#2pt height#1pt \kern#1pt
 \vrule width.#2pt}\hrule height.#2pt}}}}
\def\square{%
  \mathop{\mathchoice{\sqr{12}{15}}{\sqr{9}{12}}{\sqr{6.3}{9}}{\sqr{4.5}{9}}}}

%%%%%%%%% jtl macros
%%%%%%%%%%%%
\newcommand{\fft}[2]{{\frac{#1}{#2}}}
\newcommand{\ft}[2]{{\textstyle{\frac{#1}{#2}}}}
\def\jsquare{\mathop{\mathchoice{\sqr{8}{32}}{\sqr{8}{32}}
{\sqr{6.3}{9}}{\sqr{4.5}{9}}}}

\newcommand{\wn}{\mathfrak{w}}

%%%%%%%%% paper specific macros
%%%%%%%%%%%%

\def\a{\alpha}
\def\b{\beta}
\def\l{\lambda}
\def\w{\omega}
\def\dd{\delta}
\def\DD{\Delta}
\def\r{\rho}
\def\c{\chi}
\newcommand{\qq}{\mathfrak{q}}
\newcommand{\ww}{\mathfrak{w}}
\def\p{\phi}
\def\k{\kappa}
\def\hr{\hat{r}}
\def\om{\Omega}
\def\e{\epsilon}
\def\bm{\bar{\mu}}
\def\tc{\tilde{\calc}}
\def\tz{\tilde{z}}
\def\hz{\hat{z}}
\def\t{\tau}
\def\ra{\Rightarrow}
\def\lgb{\lambda_{\rm GB}}
\catcode`\@=12

%%%
%%%%%% text starts here
%%%%%%%%%

\title{Universality of non-equilibrium dynamics of CFTs from holography}

\author{Alex Buchel}
\email{abuchel@perimeterinstitute.ca}
\affiliation{Department of Applied Mathematics, University of Western
Ontario, London, Ontario N6A 5B7, Canada}
\affiliation{Perimeter Institute for Theoretical Physics, Waterloo, Ontario N2L 2Y5,
Canada}
\author{Stephen R. Green}
\email{sgreen@perimeterinstitute.ca}
\affiliation{Perimeter Institute for Theoretical Physics, Waterloo, Ontario N2L 2Y5,
Canada}
\author{Luis Lehner}
\email{llehner@perimeterinstitute.ca}
\affiliation{Perimeter Institute for Theoretical Physics, Waterloo, Ontario N2L 2Y5,
Canada}
\author{Steven L. Liebling}
\email{steve.liebling@liu.edu}
\affiliation{Department of Physics, Long Island University, Brookville, NY 11548, U.S.A}

%\date\today
%\date{October 20, 2014}

\begin{abstract}
  Motivated by a low-energy effective description of gauge
  theory/string theory duality, we conjecture that the dynamics of
  $SO(4)$-invariant states in a large class of four-dimensional
  conformal gauge theories on $S^3$ with non-equal central charges
  $c\ne a$ are universal on time scales $t_{\rm universal}\propto
  c\ (\cale-\cale_{\rm vacuum})^{-1}$, in the limit where the energy
  $\cale\to \cale_{\rm vacuum}$. We show that low-energy excitations
  in $c\ne a$ CFTs do not thermalize in this limit.  The holographic
  universality conjecture then implies that within the Einstein-scalar
  field system (dual to theories with $c=a$), $AdS_5$ is stable to
  spherically symmetric perturbations against formation of trapped
  surfaces within time scales $t_{\rm universal}$.
\end{abstract}

\maketitle

%%%%%%%%%%%%%%%%%%%%%%%%%%%%
{\it Holographic universality conjecture.---}
%%%%%%%%%%%%%%%%%%%%%%%%%%%%
Over the years the holographic gauge/gravity correspondence has
developed into a powerful tool to probe the physics of strongly
interacting quantum systems by mapping them into problems in classical
gravity~\cite{juan}. The quintessential example is the correspondence
between supersymmetric $\caln=4$ $SU(N)$ Yang-Mills theory and type
IIB string theory in $AdS_5\times S_5$.  In the planar limit, $N\to
\infty$ and $g_{\rm YM}^2\to 0$ with $g_{\rm YM}^2 N$ kept fixed,
quantum corrections on the string theory side can be
neglected. Furthermore, for large 't~Hooft coupling, $g_{\rm YM}^2
N\to \infty $, the holographic dual is captured by classical Einstein
gravity in a five-dimensional $AdS$ spacetime. It is well understood
that successive $\calo((g_{\rm YM}^2 N)^{-3/2})$ 't~Hooft coupling
corrections translate into higher derivative $\calo((\a')^3)$
corrections, while non-planar $1/N$ corrections correspond to $g_s$
string loop corrections.

Other examples of holographic correspondence for four-dimensional superconformal gauge theories
involve $AdS_5\times X_5$ string theory, where $X_5$ is a general Sasaki-Einstein manifold
\cite{adscft}. Once again, a Kaluza-Klein reduction on $X_5$ 
(in the planar limit and for large 't~Hooft coupling) 
produces a sector of classical Einstein gravity in $AdS_5$,
\begin{equation}
\begin{split}
S=&\frac{1}{2\ell_p^3}\int_{\calm_{5}}d^{5}\xi \sqrt{-g} 
\biggl(\frac{12}{L^2}+R+\call_{\rm matter}\biggr)\,,
\end{split}
\label{eq:cisa}
\end{equation}
where $\call_{\rm matter}$ is a Lagrangian for the gravitational bulk matter sector which encodes 
the spectrum of operators of the dual CFT with small anomalous dimensions.  

A common feature of strongly coupled 
gauge theories with gravity dual \eqref{eq:cisa} 
is the equality  between the two central charges $c$ and $a$ parameterizing the 
conformal anomaly of a four-dimensional CFT in a curved spacetime  $\calm_4$ \cite{birrell}.  More generally, the conformal anomaly takes the form
\begin{equation}
\begin{split}
&\langle T^\mu{}_\mu\rangle_{\rm CFT} =\frac{c}{16\pi^2}
I_4-\frac{a}{16\pi^2} E_4\,,\\
&E_4= r_{\mu\nu\rho\lambda}r^{\mu\nu\rho\lambda}-4
r_{\mu\nu}r^{\mu\nu}+r^2 \,,\\
& I_4=
r_{\mu\nu\rho\lambda}r^{\mu\nu\rho\lambda}-2  r_{\mu\nu}r^{\mu\nu}
+\frac 13r^2\,,
\end{split}
\label{eq:anomaly}
\end{equation}
where $E_4$ and $I_4$ correspond to the four-dimensional Euler density and the square of the Weyl curvature
of $\calm_4=\del\calm_5$. However, for the action \eqref{eq:cisa}, both central charges reduce to \cite{Balasubramanian:1999re}
\begin{equation}
c=a=\frac{\pi^2  L^3}{\ell_p^3}\,,
\label{eq:cdef}
\end{equation} 
where $L$ is the asymptotic $AdS_5$ radius of curvature, and $\ell_p$ is a five-dimensional Planck length. 

Superconformal gauge theories with $c\ne a$ can also be described in a holographic framework,
by supplementing the effective action \eqref{eq:cisa} with a five-dimensional Gauss-Bonnet (GB)
term  \cite{Banerjee:2014oaa},
\begin{equation}
\begin{split}
S=&\frac{1}{2\ell_p^3}\int_{\calm_{5}}d^{5}z \sqrt{-g} 
\biggl(\frac{12}{L^2}+R+\call_{\rm matter}\\
&+
\frac{\lgb}{2} L^2\left(R^2-4 R_{\mu\nu}R^{\mu\nu}
+R_{\mu\nu\r\sigma}R^{\mu\nu\rho\sigma}\right) 
\biggr)\,,
\end{split}
\label{eq:aisnotc}
\end{equation}
with the identifications (causality constrains 
$-\frac{7}{36}\le \lgb\le \frac{9}{100}$ \cite{Buchel:2009tt}
)
\begin{equation}
\begin{split}
&c=\frac{\pi^2 \tilde{L}^3}{\ell_p^3}\left(1-2\frac{\lgb}{\b^2}\right)\,,\qquad 
a=\frac{\pi^2 \tilde{L}^3}{\ell_p^3}\left(1-6\frac{\lgb}{\b^2}\right)\,,\\
&\tilde{L}\equiv \b L\,,\qquad \b^2\equiv \frac 12 +\frac 12 \sqrt{1-4\lgb}\,.
\end{split}
\label{eq:defca}
\end{equation}

The vacuum state of a dual CFT  is described by the solution of the effective 
action \eqref{eq:aisnotc} with the matter sector turned off,
\begin{equation}
|0\rangle\bigg|_{\rm CFT}\qquad \Longleftrightarrow\qquad \call_{\rm matter}=0\,.
\label{eq:vacuum}
\end{equation}
The dual to the vacuum state of a CFT on a three-sphere $S^3$ is then global $AdS_5$,
 \begin{equation}
ds^2 = \frac{L^2\b^2}{\cos^2 x} \left(
                                     - dt^2
                                     +{dx^2}
                                     +\sin^2x \, d\Omega^2_{3}
                                        \right) \, ,
\label{eq:adsmetric}
\end{equation}
where $d\Omega_3^2$ is the metric of $S^3$.  Following holographic renormalization 
of GB gravity developed in \cite{Liu:2008zf,Banerjee:2014oaa}, we find that the 
vacuum energy (the mass) of \eqref{eq:adsmetric}, or  the 
Casimir energy from the boundary CFT perspective, is 
\begin{equation}
\cale_{\rm vacuum}=\frac{3a}{4\tilde{L}}\,.
\label{eq:casimir}
\end{equation}

The vacuum of a CFT is static. By contrast, a generic non-equilibrium
state $|\xi\rangle$ of a CFT {\it evolves} with time. In a dual
gravitational description this evolution corresponds to the dynamics
of the coupled matter-gravity sector in \eqref{eq:aisnotc}, with the
Dirichlet boundary conditions on the non-normalizable components of
the excited bulk matter fields in $\call_{\rm matter}$. Although
\[
\frac{d}{dt} |\xi\rangle\ \ne 0\,,
\]
the fact that a CFT represents a closed system implies that the energy
of the state, $\cale_{\xi}$, is conserved,
\begin{equation}
\frac{d}{dt} \cale_{\xi}=0\,.
\label{eq:energy}
\end{equation}

We will argue in the following sections using the multiscale analysis
of~\cite{Balasubramanian:2014cja} that in the limit of small
${(\cale_{\xi}-\cale_{\rm vacuum})}/{\cale_{\rm vacuum}}$, the
dynamics of the bulk are well described by the leading order
gravitational self-interaction of the matter up to a time $t_{\rm
  universal}\propto c\ (\cale_{\xi}-\cale_{\rm vacuum})^{-1}$, where $c$ 
is the CFT central charge.  It is
therefore reasonable to assume that on this time scale, the physics of
the strongly coupled dual CFT is also described by this approximation.
Moreover, to leading order the gravitational self-interaction is
independent of $\lgb$. We are thus led to the following {\em holographic universality conjecture}:
\\

{\it Consider an $SO(4)$ invariant state $|\xi\rangle$ of a CFT. The
  dynamics of the CFTs are universal in the limit
  ${(\cale_{\xi}-\cale_{\rm vacuum})}/{\cale_{\rm vacuum}}\to 0$:
  Apart from a simple rescaling, they are insensitive to the central
  charge difference $(c-a)/c$ up to time scales $t_{\rm universal}$.
}
\\

An important consequence of the above is the stability of $AdS_5$ on time scales 
$t_{\rm universal}$:

{\it 
CFTs with non-equal central charges $(c-a)/c\ne 0$, which 
allow for an effective holographic description \eqref{eq:aisnotc},
cannot thermalize unless
\begin{equation}
\cale_\xi-\cale_{\rm vacuum} \ge \frac{3c}{\tilde{L}}
\begin{cases}
\frac{1-\b^2}{2\b^2-1}\,, &{\rm if}\ \lgb>0\,,\\
{(\b^2-1)}{(2\b^2-1)}\,, &{\rm if}\ \lgb<0\,.
\end{cases}
\label{eq:defde}
\end{equation}
Thus, under the assumption of universality, spherically symmetric
Einstein gravity plus matter in $AdS_5$, which is dual to CFTs with
$c=a$, cannot thermalize within time scales of order $t_{\rm
  universal}$---the instability \cite{box}
(in the sense of \cite{Bizon:2011gg}) can arise only on time scales
longer than $t_{\rm universal}$ (as indicated by the plots
in~\cite{Buchel:2013uba}) .  }

The conformal gauge theories discussed in this work are 
somewhat restrictive. 
First, they must have a holographic dual within the supergravity approximation of type IIB string theory. 
Second, the higher derivative gravitational 
corrections (required to produce the central charge difference, $c\ne a$) 
are assembled into the GB combination (this is necessary to ensure that 
all the bulk equations of motion are of the 
second order). Finally, we impose some 
technical restrictions on the possible form of $\call_{\rm matter}$ in 
\eqref{eq:aisnotc}.  It would be interesting to relax these constraints, and
also to consider generalizing the discussion to holographic quantum theories with hyperscaling
violation \cite{Dong:2012se}.

%-----------------------
{\it Evidence for the conjecture.---}
%\label{sec:evidence}
%-----------------------
The following assumption is vital for the holographic universality
conjecture:

{\it The leading order self-gravitational interaction of $\call_{\rm
    matter}$ in \eqref{eq:aisnotc} correctly captures the physics of a
  holographically dual CFT at low energy.}

The matter sector in \eqref{eq:aisnotc} represented by $\call_{\rm
  matter}$ can be very complicated.  Since we are interested in
non-equilibrium dynamics of $SO(4)$ invariant states $|\xi\rangle$
with $\cale_{\xi}$ close to the vacuum energy, we take the approximation
\begin{equation}
\call_{\rm matter}=\sum_{i}\ \left(- 3 \del_\mu\phi_i\del^\mu\phi_i - 3 m_i^2 \phi_i^2\right)+\calo(\phi^6)\,,
\label{eq:lmatter}
\end{equation} 
where the summation runs over the set of bulk scalar fields
$\{\phi_i\}$, dual to the spectrum of operators $\{\calo_i\}$ of
dimensions $\{\Delta_i\}$ in a boundary CFT,
\begin{equation}
\Delta_i (4-\Delta_i) = - m_i^2 \tilde{L}^2\,.
\label{eq:dmcorre}
\end{equation}
We study the evolution of a state in a CFT, and, as a result, we impose Dirichlet boundary conditions on non-normalizable coefficients of 
 $\{\phi_i\}$.
Note that we restricted possible 
potential terms in \eqref{eq:lmatter}, 
\ie, we assumed that the matter sector is well approximated 
by essentially free bulk scalars. Such an approximation implies that 
leading nonlinearities in the bulk dynamics come from the 
interactions through gravity (minimal coupling), rather than from 
non-linear interactions within the matter sector itself. Additionally, 
we assume that $\Delta_i >2 $.

To proceed, we write the $5$-dimensional metric describing an
asymptotically AdS spacetime with $SO(4)$ symmetry in the form
\begin{equation}
ds^2 = \frac{L^2\b^2}{\cos^2 x} \left(
                                     -Ae^{-2\delta} dt^2
                                     +\frac{dx^2}{A}
                                     +\sin^2x \, d\Omega^2_{3}
                                        \right) \, ,
\label{eq:metric}
\end{equation}
where $A(x,t)$ and $\delta(x,t)$ are scalar functions. We also take
the scalar fields to be functions of $(t,x)$ only:
$\phi_i=\phi_i(t,x)$.
\begin{widetext}
  With the above ansatz, we obtain the equations of motion,
\begin{eqnarray}
\label{eq:constx1}0&=&\Box \phi_i+\Delta_i(\Delta_i-4)\phi_i\,,\\
\label{eq:constx2}A_{,x}       &=&    \frac{1}{\cos x(\b^2 \sin^2 x +2\lgb (\cos^2 x-A))}\biggl(
2 \sin x (\b^2(1+\sin^2x) (\b^2-A)-\b^2(\b^2-1)\cos^2 x-2\lgb A(\cos^2 x-A))
\biggr)\nonumber \\
&&     - \frac{\b^2 \sin^3 x \cos x }{A(\b^2 \sin^2 x +2\lgb (\cos^2 x-A))}
 \sum_i \biggl(
e^{2\dd} (\del_t \phi_i)^2 +A^2 (\del_x\phi_i)^2
+\frac{A}{ \cos^{2} x}\ \Delta_i(\Delta_i-4)\ \phi_i^2 
\biggr)\,,\\
\label{eq:constx3}\delta_{,x}  &=&   - \frac{\b^2 \sin^3 x  \cos x }{A^2 (\b^2 \sin^2 x +2\lgb (\cos^2 x-A))} 
\sum_i \left( e^{2\dd } (\del_t \phi_i)^2
 +A^2 (\del_x\phi_i)^2\right)\,,
\end{eqnarray}
where $\Box$ is computed with \eqref{eq:metric},
together with one constraint equation,
\begin{equation}
A_{,t} +  \frac{2\b^2 \sin^3 x \cos x  A }{\b^2 \sin^2 x +2\lgb (\cos^2 x-A)}  \sum_i
\del_t \phi_i\del_x \phi_i =0\,.
\label{eq:consteq}
\end{equation}
\end{widetext}
There is an additional second order equation, which is however redundant due to the $SO(4)$ symmetry. 
Notice that the $AdS_5$ solution \eqref{eq:adsmetric} is recovered  with
\begin{equation}
\phi_i=\delta=0\,,\qquad A=1\,.
\label{eq:ads5}
\end{equation}

We are interested in smooth solutions of \eqref{eq:constx1}--\eqref{eq:constx3}, subject to the following 
boundary conditions.
At the origin, regularity implies
\begin{equation}
\begin{split}
\phi(t,x)  = & \phi_0(t) + {\cal O}(x^2)\,,  \\
A(t,x)       = &   1 + {\cal O}(x^2)\,,  \\
\delta(t,x)  = &  \delta_0(t) + {\cal O}(x^2)\,.
\end{split}
\label{eq:ir}
\end{equation}
At the outer boundary $x=\pi/2$ we introduce $\rho \equiv \pi/2-x$ so that 
\begin{equation}
\begin{split}
\phi_i(t,\rho)  = & \rho^{\Delta_i}\left(\phi_{\Delta_i}(t) + {\cal O}(\rho^2)\right)\,,  \\
A(t,\rho)       = & 
 1 - M \r^4 +  {\cal O}(\rho^{6})+  {\cal O}(\rho^{2\min_i{(\Delta_i)}})\,,  \\
\delta(t,\rho)  = &  0 + {\cal O}(\rho^{2 \min_i{(\Delta_i)}})\,.
\end{split}
\label{eq:uv}
\end{equation}
\begin{widetext}
The parameter $M$ in \eqref{eq:uv} is related to the conserved energy $\cale_{\xi}$.
Indeed, as described in \cite{Buchel:2012uh} it is convenient to introduce the mass-aspect 
function $\calm(t,x)$ as 
\begin{equation}
A(t,x)=1-\frac{1}{2 \lgb} \biggl((2 \lgb-\b^2) \sin^2 x
+\biggl(4 \lgb (\b^2-2 \lgb) \calm(t,x) \cos^4 x+(2 \lgb-\b^2)^2 \cos^4 x
-\b^4 (1-4 \lgb) \cos(2 x)\biggr)^{1/2}\biggr)\,.
\label{eq:massaspect}
\end{equation}
Using \eqref{eq:constx2} we conclude that 
\begin{equation}
\calm(t,x)=\frac{1}{2\b^2-1}\ \int_0^x dz \frac{\tan^3 z}{A(t,z)}\  
\sum_i \biggl[e^{2\dd} (\del_t \phi_i)^2
 +A^2 (\del_x\phi_i)^2+\frac{A}{ \cos^{2} x}\ \Delta_i(\Delta_i-4)\ \phi_i^2 \biggr] \, .
\label{eq:defcalm}
\end{equation}
\end{widetext}
Furthermore, 
\begin{equation}
M=\calm(t,x)\bigg|_{x=\frac \pi 2}\,.
\label{eq:compM}
\end{equation}
Using the machinery of holographic renormalization \cite{Liu:2008zf,Banerjee:2014oaa},
 we compute the energy of $|\xi\rangle$
\begin{equation}
\cale_\xi=\frac{3c}{4L \b} \biggl(\frac{\b^2-6\lgb}{\b^2-2\lgb}+4 M\biggr)
=\frac{3c}{4\tilde{L}} \biggl(\frac{a}{c}+4 M\biggr)\,.
\label{eq:energyxi}
\end{equation}

We will show now that to leading order in the backreaction, the dynamics of 
\eqref{eq:constx1}--\eqref{eq:constx3} is universal---apart from a simple rescaling, it is insensitive to 
$\lgb$.
We apply the ``Two Time Framework'' (TTF)
introduced in \cite{Balasubramanian:2014cja} to this system: some number of
possibly massive scalar fields coupled to GB gravity in $AdS_5$.
We account for the backreaction by 
introducing a parameter $\e$ and the associated slow time 
\begin{equation}
\t \equiv  s_1 \e^2 t \,.
\label{eq:slow}
\end{equation}
We expand the fields in terms of both the fast time $t$ and slow time $\t$ as 
\begin{eqnarray}
\phi_i &=& \e \biggl(\phi_{i,(1)}(t,\t,x)+s_2\ \e^2 \phi_{i,(3)}(t,\t,x)+\calo(\e^4) \biggr)\,,\nonumber\\
\label{eq:pertexp}A&=&1+s_2\ \e^2 A_{(2)}(t,\t,x)+\calo(\e^4)\,,\\
\dd&=&s_2\ \e^2 \dd_{(2)}(t,\t,x)+\calo(\e^4)\,,\nonumber
\end{eqnarray}
where $s_i$ are $\lgb$-dependent constants. From \eqref{eq:constx1} we find
\begin{eqnarray}\label{eq:phi1}
  \partial_t^2{\phi}_{i,(1)} &=& \phi_{i,(1)}'' + \frac{3}{\sin x\cos x}\phi_{i,(1)}'
-\frac{\Delta_i(\Delta_i-4)}{\cos^2 x}\phi_{i,(1)}\nonumber\\
 &\equiv& -L_{i}\phi_{i,(1)}\,.
\end{eqnarray}
The operator $L_i$ has eigenvalues $\omega_{i,j}^2=(2j+\Delta_i)^2$ ($j=0,1,2,\ldots$)
and eigenvectors $e_{i,j}(x)$ (``oscillons'').  Explicitly,
\begin{equation}
  e_{i,j}(x)=d_{i,j}\cos^{\Delta_i} x\ _2 F_1\left(-j,\Delta_i+j;2; \sin^2 x\right)\,,
  \label{eq:adef}
\end{equation}
with $d_{i,j}$ the normalization constants.  The oscillons form
an orthonormal basis under the inner product
\begin{equation}
  (f,g)=\int_0^{\pi/2}f(x)g(x)\tan^3x\,\mathrm{d}x\,.
\end{equation}
 The general real solution to \eqref{eq:phi1} is
\begin{equation}\label{eq:phi1sol}
  \phi_{i,(1)}(t,\tau,x) = \sum_{j=0}^\infty \left(A_{i,j}(\tau)e^{-i\omega_{i,j}t}+\bar{A}_{i,j}
(\tau)e^{i\omega_{i,j} t}\right)e_{i,j}(x)\,,
\end{equation}
where $A_{i,j}(\tau)$ are arbitrary functions of $\tau$, to be
determined later.

\begin{widetext}
At $O(\epsilon^2)$ the constraints \eqref{eq:constx2}--\eqref{eq:constx3} have solutions
\begin{eqnarray}
\label{eq:A2} A_{(2)}(x) &=& -\frac{1}{s_2(2\b^2-1)}\frac{\cos^4x}{\sin^2 x}\int_0^x\
\sum_i \biggl(|\del_y\phi_{i,(1)}(y)|^2
+|\del_t\phi_{i,(1)}(y)|^2+\frac{\Delta_i(\Delta_i-4)}{\cos^2 x} \phi_{i,(1)}(y)^2\biggr)\tan^3y \,\mathrm{d}y\,,\\
 \delta_{(2)}(x) &=& \frac{1}{s_2(2\b^2-1)}\int_x^{\pi/2}\ \sum_i \left(|\del_y \phi_{i,(1)}(y)|^2
+|\del_t \phi_{i,(1)}(y)|^2\right)\sin y\cos y \,\mathrm{d}y\,.
\label{eq:d2}
\end{eqnarray}
Note that we are using the gauge with $\dd_{(2)}(\pi/2)=0$.
\end{widetext}

Finally, at $O(\epsilon^3)$ we obtain the equations for $\phi_{i,(3)}$,
\begin{equation}
 \partial_t^2\phi_{i,(3)} + L_i\phi_{i,(3)}
+ \frac{2s_1}{s_2}\partial_t\partial_\tau\phi_{i,(1)} =S_{i,(3)}(t,\tau,x)\,,
\label{eq:phi3}
\end{equation} 
where the source term is 
\begin{align}
  S_{i,(3)}={}&\del_t(A_{(2)}-\delta_{(2)})\del_t\phi_{i,(1)}-2(A_{(2)}-\dd_{(2)}) L_i\phi_{i,(1)}\nonumber\\
  &+(A_{(2)}'-\dd_2')\phi_{i,(1)}'+\frac{\Delta_i(\Delta_i-4)A_{(2)}}{\cos^2 x} \phi_{i,(1)}\,.
\label{eq:s3}
\end{align}

Note that by choosing 
\begin{equation}
s_1=s_2=\frac{1}{2\b^2-1}= \frac{1}{\sqrt{1-4\lgb}}\,,
\label{eq:sichoice}
\end{equation}
the dependence on $\lgb$ in TTF is completely factored 
out \cite{Myers:2010ru}. 
In other words, 
the rescaling \eqref{eq:sichoice} identifies the TTF equations for different $\lgb$,
\begin{equation}
\biggl(\e^2,\ \lgb\biggr)\, \Longleftrightarrow\,
\biggl(\e_{\rm eff}^2=\frac{\e^2}{\sqrt{1-4\lgb}},\ {\lgb}^{\rm eff}= 0\biggr)\,.
\label{eq:relate}
\end{equation} 
The identification \eqref{eq:relate} is the basis for our holographic universality conjecture:
since the TTF framework is valid on slow time scales, we expect universality of low-energy,
non-equilibrium dynamics of dual CFTs on time scales
\begin{equation}
t_{\rm universal}\propto \epsilon^{-2} \propto c\ (\cale_{\xi}-\cale_{\rm vacuum})^{-1}\,,\qquad \e\to 0\,.
\label{eq:tuniversal}
\end{equation}  

It is important to emphasize that the the scalar field profile must be
held fixed as the limit $\e\to 0$ is taken.  On the CFT side, this
corresponds to fixing the state as the energy is taken to $\cale_{\rm
  vacuum}$.  Were the profile allowed to vary, then backreaction could be increased by concentrating the field energy into a small region.
Indeed,
consider the initial condition prepared by a marginal operator 
with $\Delta_i=4$, of the form
\begin{equation}
\begin{split}
\phi_{{(1)}}\bigg|_{t=0}=&~0\,,\\
\del_t \phi_{(1)}\bigg|_{t=0} =& ~\w_j\ d_j\ \cos^4 x \
_2 F_1\left(-j,4+j;2; \sin^2 x\right),
\end{split}
\label{eq:initialhigh}
\end{equation}
with $d_j=2\sqrt{(j+1)(j+2)(j+3)}$.
The mass parameter $M$ corresponding to this profile is 
\begin{equation}
M=\e^2 \w_j^2=\e^2 (4+2j)^2 .
\label{eq:defmj}
\end{equation} 
We can keep $M$ fixed in the limit $\e\to 0$ if we excite the 
oscillon with index $j \sim M^{-1} \gg 1$. Following \eqref{eq:A2} 
we estimate for $x\lesssim j^{-1}$
\begin{equation}
\begin{split}
|A_{(2)}(x,t=\t=0)|&\sim \frac{\cos^4 x}{\sin^2 x}\ x^4 \w_j^2\ d_j^2\\
&\sim  \frac{M}{\e^2}\ (x\ d_j)^2 \sim \frac{M j}{\e^2}\ (x j)^{2}.
\end{split}
\label{eq:estA2j}
\end{equation}
Thus, $|\e^2 A_{(2)}|_{(t,\t)=(0,0)}$ becomes of the same order as the
leading contribution for $x\sim j^{-1}$ and the series expansion
\eqref{eq:pertexp} is inconsistent.  When we talk about the leading order
backreaction in the limit $(\cale_{\xi}-\cale_{\rm vacuum})\to 0$, we
always assume that $\e\to 0$ with the initial profile shape kept
fixed. A priori, this does
not guarantee that during the evolution $|A_{(2)}(t,\t)|$ will
continue to remain bounded. We argue in the next section that if
$|A_{(2)}(t,\t)|$ is bounded initially, it must be bounded for all
times.

{\it Universal dynamics and $AdS_5$ (in)stability.---}
Consider $c\ne a$ CFTs with dual holographic descriptions ($\b\ne 1$). Equilibrium thermal states of such models are described by the static  solution  
\begin{equation}
\begin{split}
A=&~1-\frac{1}{2 \lgb} \biggl((2 \lgb-\b^2) \sin^2 x\\
&+\biggl(4 \lgb (\b^2-2 \lgb) M \cos^4 x\\
&+(2 \lgb-\b^2)^2 \cos^4 x
-\b^4 (1-4 \lgb) \cos(2 x)\biggr)^{1/2}\biggr)\,,\\
\phi_i=&~0\,,\qquad \dd=0\,.
\end{split}
\label{eq:statbh}
\end{equation} 
It is straightforward to observe that \eqref{eq:statbh} has a regular horizon only if 
\begin{equation}
M \ge 
\begin{cases}
\frac{1-\b^2}{2\b^2-1}\,, &{\rm if}\ \lgb>0\,,\\
{(\b^2-1)}{(2\b^2-1)}\,, &{\rm if}\ \lgb<0\,.
\end{cases}
\label{eq:defm}
\end{equation}
Thus, generic non-stationary states in such CFTs cannot equilibrate in the limit 
$(\cale_{\xi}-\cale_{\rm vacuum})/\cale_{\rm vacuum}\to 0$ [cf.~eq.~\eqref{eq:defde}].

Under the assumption of holographic universality (\ie, that the
leading order self-gravity dynamics in the bulk correctly capture the
behavior of the CFT for $t<t_{\text{universal}}\propto \e^{-2}$),
these arguments imply that Einstein gravity in $AdS_5$ is stable
within time scales $t_{\rm universal}$ against scalar collapse of
generic initial data with amplitude $\propto \e$, in the limit $\e\to
0$.  Indeed, since black holes cannot form for arbitrarily small
$\epsilon$ in Gauss-Bonnet gravity \cite{Deppe:2014oua,inprog},
holographic universality implies TTF solutions cannot diverge in
finite time.  Thus, since TTF dynamics are independent of $\lgb$,
Einstein gravity in particular is stable for
$t<t_{\text{universal}}$. Of course, there is no tension with
conjectures for instability of AdS
~\cite{Anderson:2006ax,holzegeldafermos}, since collapse can still occur
over longer time scales.

{\it TTF validity within fully non-linear dynamics and String Theory.---}
The holographic universality conjecture (and the supporting evidence) 
assumes that the leading order self-gravitation faithfully captures the dynamics of 
a generic state $|\xi\rangle$ in the CFT, with a holographic dual,
in the limit $(\cale_{\xi}-\cale_{\rm vacuum})/\cale_{\rm vacuum}\to 0$. We stress that 
if this is not true, not only do the higher order terms in the perturbative 
expansion \eqref{eq:pertexp} become
important, but so do the $\calo(\phi^6)$ corrections to the scalar potential in   
\eqref{eq:lmatter}.

In particular, if the higher order terms  encoding  the mass gap in GB gravity
\eqref{eq:defm} are important, then the TTF equation for $A_{(2)}$ \eqref{eq:A2} 
might develop a singularity
in a finite slow time $\tau\to \t_{\rm singular}$,
\begin{equation}
\lim_{\t\to \t_{\rm singular}} A_{(2)}\to -\infty\,.
\label{eq:singa2}
\end{equation}
This would result in the formation of a trapped surface within time
scales $\propto \e^{-2}$; earlier than the scenario described in the
previous section.  It would be interesting to explore this further and
to connect to numerical analysis in
\cite{Jalmuzna:2011qw,Buchel:2012uh}. However, numerical results to
date (for $d=4$, see figure in \cite{Buchel:2013uba}) indicate that
$t_{\text{collapse}} > t_{\text{universal}}$, consistent with the
conjecture. 

The universality conjecture presented in this Letter has been motivated 
in the holographic framework, within the supergravity approximation to String 
Theory. Thus, it relies on the validity of the effective 
gravitational action \eqref{eq:aisnotc}. Recently, it was pointed 
out that GB effective actions arising from a consistent theory of 
gravity must include a tower of massive higher-spin states 
\cite{Camanho:2014apa}. These states can not be excited in 
$SO(4)$-symmetric dynamics discussed here. It would be interesting 
to relax the symmetry constraint and re-analyze approach to 
equilibrium of the boundary CFT. 
Note, however, that from the gravitational perspective it is natural
to expect black hole formation is most efficient in spherical settings. 

While the initial conditions for the CFT dynamical evolution
in the holographic framework 
are well described by the supergravity modes, 
the late-time dynamics, in particular the formation 
of small black holes, can require physics beyond the supergravity 
approximation. Specifically \cite{juanprivate}, 
the approach to equilibrium 
of the boundary CFT can proceed through the initial formation 
of the highly excited ball of strings in the bulk, which is
expected to be  entropically favourable compared to the corresponding 
Schwarzschild black hole \cite{Horowitz:1996nw}. 
Such a process necessitates decay of the 
supergravity modes into string states, which is 
unfortunately not well understood.

{\bf Acknowledgments:}
We would like to thank A.~Frey and G.~Kunstatter for interesting discussions 
and correspondence. We also thank J.~Maldacena and R.~Myers 
for valuable discussions
and  comments on the 
manuscript.  This work
was supported by the NSF under grant
PHY-1308621~(LIU), by NASA under grant NNX13AH01G, by NSERC through a
Discovery Grant (to A.B. and L.L.) and by CIFAR (to L.L.).
Research at Perimeter
Institute is supported through Industry Canada and by the Province of
Ontario through the Ministry of Research \& Innovation.  Computations
were performed at Sharcnet.


\begin{thebibliography}{99}
\bibitem{juanprivate} We would like to thank Juan Maldacena 
for suggesting this possibility.
\end{thebibliography}
\end{document}